# Security, Privacy and Challenges in Microservices Architecture and Cloud Computing- Survey

Hemanth Gopal
hemantg1@umbc.edu
*University of Maryland Baltimore County*

Guanqun Song
Song.2107@osu.edu
*The Ohio State University*

Ting Zhu
zhu.3445@osu.edu
*The Ohio State University*

*Abstract*—Security issues in processor architectures remain really critical since users and devices continue to share computing as well as networking resources.So, preserving data privacy in such an environment is really a critical concern.We know that there is a continuous growth in security and privacy issues that need to be addressed.Here, we have chosen a microservice architecture, which is a small or even an independent microprocess that interacts, acts, and responds to messages via lightweight technologies such as Thrift, HTTP, or REST API.

## I. INTRODUCTION

The migration of monolithic architecture as shown in Figure [1] to the cloud has proven to be a significant challenge. Microservices have been adopted as a natural option in the replacement of monolithic systems, according to the findings of this work. The fundamental concern is how its design has been implemented, as well as security and privacy considerations in a cloud computing environment. The fact that microservices are increasingly being used as a solution for cloud-based systems inspired this collection. Cloud computing, like utility-based systems like electricity, water, and sewage, provides a centralized pool of configurable computing resources and computer outsourcing methods that enable diverse computing services to different persons [3].

Java, C, C/C++, Ruby, and Python are now the most widely used server-side development programming languages in academic and professional settings, owing to their ability to provide linear and concrete abstractions that reduce program complexity by splitting them down into single modules. This is the primary goal, capability, and purpose for creating unique executable artefacts, sometimes known as monoliths, whose modularization abstractions rely on the sharing of the same machine's resources [5]. Despite being a simple model of use, the limits are clear and do not stop at scalability; they extend well beyond and cross the most critical part of today's society, security.[10][11] The Microservices method is a first manifestation of SOA that emerged after the introduction of DevOps and is rapidly gaining popularity for developing constantly deployed applications. Netflix made one of the most well-known transitions from monolithic to microservices in 2010, when they began to use AWS Amazon to host its application and services in more than 100 granular services, rather than a single web-app (.war). Other companies, such as eBay, now use it. [6][13]. Microservices' success is based on where Monolithic fails, such as problematic issues such as the need to rebuild an entire development just to change a small constraint or check, or the risk factor that today's large vertical applications may have major failures or bugs that jeopardize the entire application's purpose. Monolithic development is a full stack or vertical alignment of an item, which requires more resources. However, for tiny and basic point applications, it is not only a possible solution, but it is the correct one.[4]

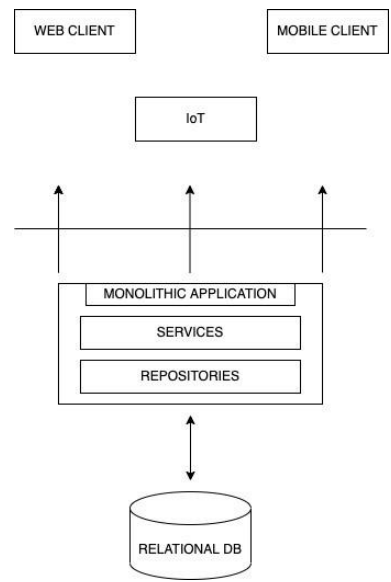

Fig. 1. Monolithic Architecture

Microservices as shown in Figure [2] is an architectural paradigm that combines Service-Oriented Architecture and the classic Unix principle of "do one thing and do it well." Microservices are designed to be light, flexible, and simple to use, and they align with modern software engineering trends like Agile development, Domain Driven Design (DDD), cloud, containerization, and virtualization. Organizational alignment,

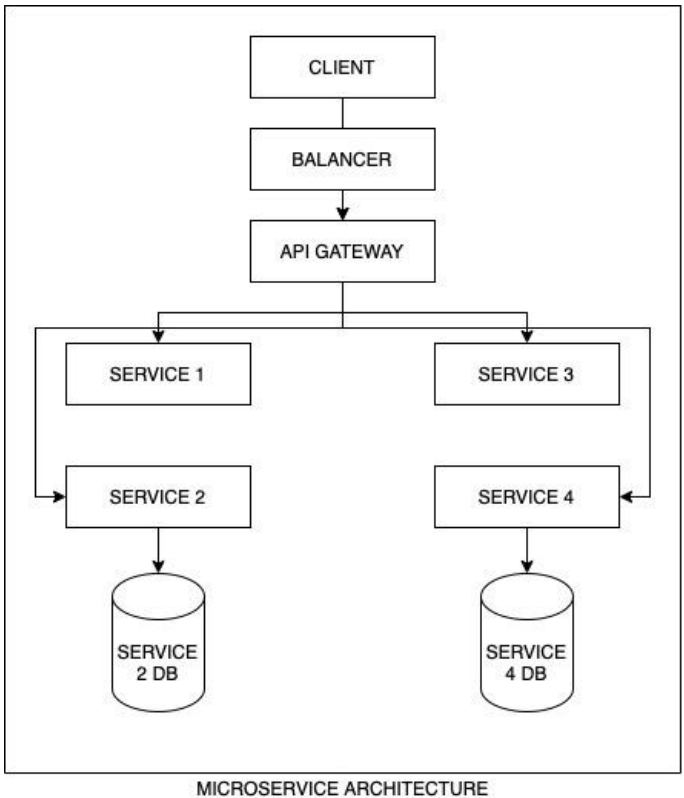

Fig. 2. Microservice Architecture

faster and more frequent software releases, independent scalability of components, and overall faster technology adoption are the most commonly mentioned benefits of microservice architecture. When compared to traditional non-distributed solutions, the disadvantages include the requirement for many design choices, the complexity of testing and monitoring, and operational overhead.[3][15]

One common criticism leveled at the microservices architecture is that it raises the application's security risk by widening the danger surface. This is true in the sense that as the number of microservices that expose functionality to external customers grows, we must safeguard each service from them. We may have most of these functions inside internal programs that are not available to external consumers in a monolithic application.[1]

## II. DETAILED ANALYSIS OF MICROSERVICE ARCHITECTURE

### A. Properties

*1) Scaling Microservices:* The major one is the fast scalability procedure, which is one of these features that makes using this method a primary alternative for developers. This approach allows Microservices to scale independently by increasing CPU or memory on the X axis, also known as vertical scaling or scaleup, or by sharding on the Z axis, also known as horizontal scaling or scale out.[4]

*2) Indepenednt Update Feature:* Each service can be set up independently of the others. A developer can simply change the availability of a service without the requirement for collaboration with other development teams or members. It also serves as a catalyst for the development of continuous integration and delivery.[4]

*3) Simple Maintenance:* Microservice code is limited to one capability, making it easier to comprehend than monolithic design. Small portions of code can be easily loaded by IDEs, making the build lighter. Working with smaller code bases speeds up development and gives programmers a better sense of the code side effects they're changing or producing.[4]

*4) Heterogeneity and Multilingualism:* Programmers are free to use whichever programming languages they like for their service, as long as they stay within the service's parameters. This enables the service to be rewritten utilizing cutting-edge technologies, as well as the freedom to select the technology, tools, and structure.[4]

*5) Failure and Isolated resource:* In a monolithic design, a badly developed service with issues such as a lack of memory or an unconnected database connection would cause performance issues or even entire program failure; however, in a Microservices architecture scenario, only that service is affected. Microservices isolate problems and lessen the impact of a breakdown on an application. Failures in well-designed microservices are isolated in one service and do not spread to the rest of the system, leaving the end user with no indication of weakness.[4]

*6) Improved communication between teams:* A Full-Stack team usually creates a microservice. All people associated with a domain work together in a single team, which greatly increases team communication. It becomes advantageous because they have similar end goals, provide a consistent schedule, and, probably most importantly, provide service as a team product.[4]

### B. Distributed Systems

Many authors use the fallacy of comparing microservices to monolithic applications while discussing microservices. The reality is more complicated. In fact, a "monolithic" program may be very modular on the inside, consisting of a large number of components and libraries from several vendors, with some components (such as a database) dispersed throughout the network. Whether API requests are done locally or via the network, the difficulties of decomposition, concern separation, and developing and describing APIs will be identical.[1]

Microservices are extremely modular, distributed systems that can be reused thanks to a network-exposed API. Microservices thus inherit the benefits and drawbacks of both distributed systems and web services. While distributed systems have numerous advantages, such as on-demand scalability and the acceptance of heterogeneity, they also have disadvantages. In the first place, developing distributed systems is more difficult.

Furthermore, distributed systems are more difficult to monitor, test, and debug than non-distributed systems. Main-taining data consistency and replication among nodes, node identification and discovery owing to continually changing network architecture, and dealing with unstable networks are all well-known issues. The following are some key aspects that differentiate distributed systems from monolithic systems:

1) Individual nodes are unaware of the overall system state
2) Individual nodes make decisions based on locally available information;
3) A node failure should not influence other nodes

These requirements are enforced by the microservice architecture.

*C. Microservices - In the context on other technologies being currently used*

Unikernels is a technology that is closely tied to the microservice paradigm. Unikernels are single-purpose OS kernels that run on a hypervisor directly. Multiple unikernels executing different functions in a distributed manner would make up a whole system. The unikernels universe fits well with the microservice design ideas of small, loosely connected single-purpose components. Unikernels promise to give excellent virtual machine security and isolation while remaining lightweight like containers.[1][7]

Some programming languages, such as Erlang and its descendant Elixir, are built to support distributed computing. Erlang is a programming language that is actor-based. Everything in Erlang is a process, and the only way to interact with them is through message passing. Encapsulation/process isolation, parallelism, fault detection, location transparency, and dynamic code upgrading are all aspects of the Erlang language [16, p.29]. Erlang features sophisticated monitoring capabilities and is utilized in a number of high-volume applications, like the messaging service WhatsApp. Erlang-based systems are naturally microservice-like. The essential microservice ideas can be rediscovered in numerous technologies, as the aforementioned examples demonstrate.

## III. SECURITY ISSUES

There are several security issues that are concerning the microservices architecture as shown in the Figure [3].

*A. Container Issues*

Microservices development and deployment become more agile when containers are used in the cloud. Using the same kernel as the host of many containers, on the other hand, could allow attackers to obtain unauthorized access to a container without the users' knowledge. Apart from the kernel vulnerability, there are various additional potential container concerns, including Denial of Service attacks, Container Escapes, Poisoned Images, and Secret compromise. To prevent security attacks on Microservices, it is vital to secure containers.[6]

In containers, there are two types of adversaries: direct and indirect. Direct adversaries can damage or change network and system files, as well as the fundamental services inside containers. Direct adversaries could target several system components locally or remotely.

Indirect adversaries have the same capabilities as direct adversaries, but they use the containers ecosystem to reach the software environment, such as code and image repositories. The entire deployment tool chain is included in the containers attack surface. Image conception, image distribution, automated builds, image signature, host configuration, and third-party components are all part of the deployment tool chain. The containers hub (e.g., Docker Hub) and other registries in the containers ecosystem are also sources of vulnerabilities in image distribution. Furthermore, the configuration of the container image distribution adds numerous external processes, increasing the attack surface globally. [4]

*B. Data Issues*

Because of their fine granularity, microservices require more complicated communication. As a result, there is a risk that not only message data will be intercepted, but that competitors will be able to infer company operations from message data. Microservices suffer from privacy difficulties in addition to message transport because they are frequently deployed in cloud settings, and cloud consumers are concerned that their stored information could be hacked or misused. Customers may be more skeptical of their personal information if microservices are deployed in multiple dispersed containers. The question of how to keep transmitted messages from leaking is still open.[2]

Data security is more likely to discover security vulnerabilities and harmful activity within the services themselves. The Microservices provider must provide minimum data security capabilities, such as an encryption schema to protect all data in the shared storage environment, stringent access controls to prevent unauthorized access to the data, and safe storage for scheduled backup data, to ensure data confidentiality, integrity, and availability.[4]

With the growing popularity of Microservices-based fog applications hosted in the cloud, privacy and confidentiality are becoming increasingly important. Tenants using Microservices-based fog applications (e.g., an individual, a business, a government agency, or any other entity) are concerned about the security of their personal information.[4]

*C. Permission Issues*

Microservices are commonly used in a distributed computing environment, which can lead to security vulnerabilities due to poor access control measures. This section discusses permission security issues in Microservices, such as authentication and authorisation.[4] Because a single service controlled by an attacker might maliciously impact other services, a microservices architecture should verify the authenticity of each service. When a service receives a message, it must also determine whether the message is genuine and whether the sender service has valid authority. According to Gegick and Barnum, a subject should only be granted the bare minimum of rights when requesting access to a resource, and those rights should be granted for the shortest time possible. Giving a user permissions beyond what is required can allow that user to obtain or change information in an unauthorized manner. As a result, delegating access rights carefully can prevent intruders from causing damage to a system.[2]

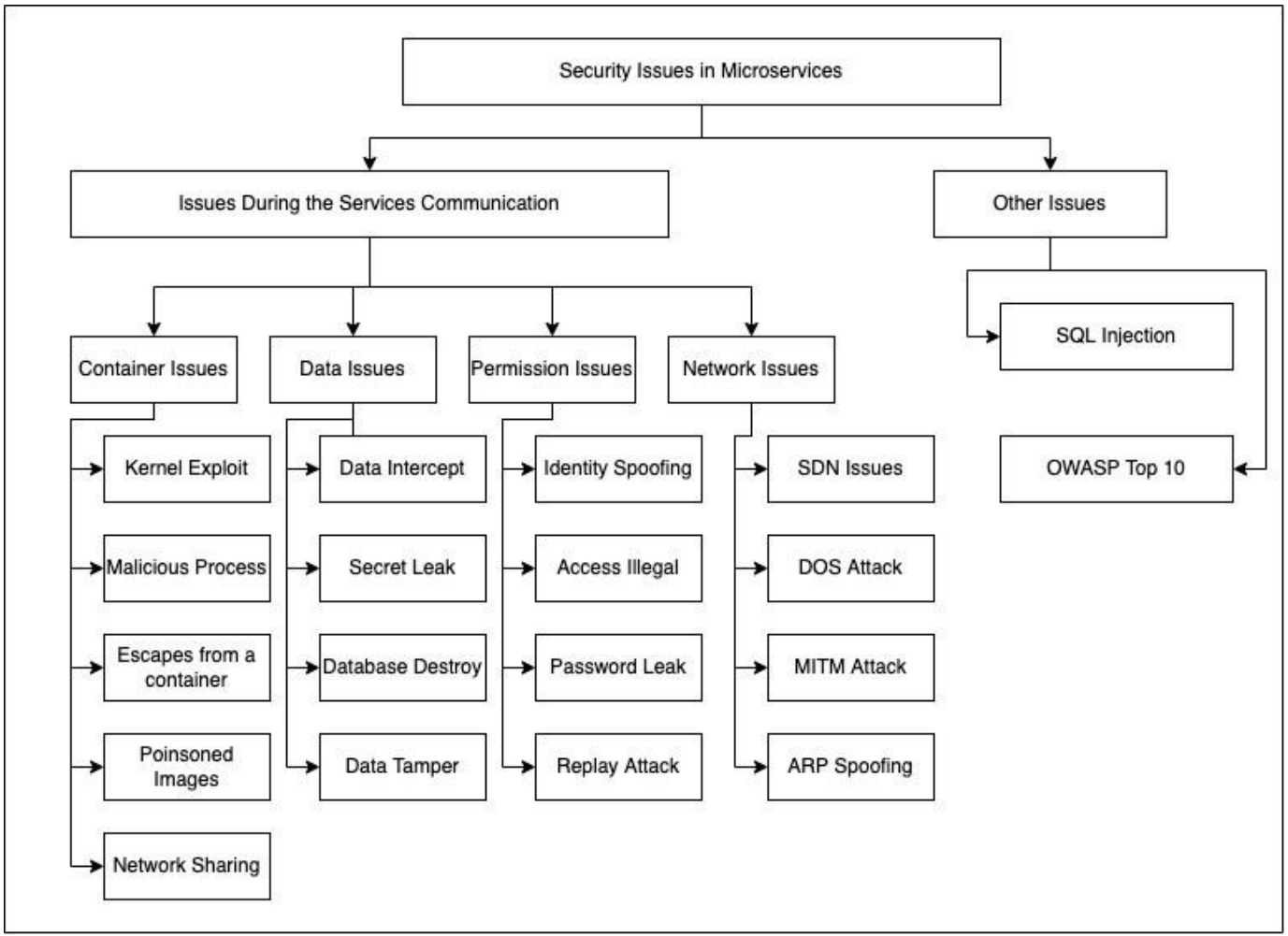

Fig. 3. Security Issues in Microservices Architecture

### D. Network Issues

With the growth of the Internet, there have been many unauthorized visits, malicious assaults, and other network difficulties that have harmed businesses and individuals. 80 As a result, network security is gradually becoming a critical issue that we cannot overlook. Various solutions have been introduced to increase network security. 81-85 Furthermore, because services communicate more frequently in Microservices, network security is crucial. The entire Microservices framework would be crippled if a network attack caused a node delay or outage. Only secure networks can assure the communication of secure services. We will discuss network dangers and successful solutions in this part.[2]

## IV. SECURITY ON CLOUD COMPUTING FOR MICROSERVICES

### A. Security for Cloud computing

Switching from a monolithic or centralized to a decentralized architecture necessitates caution. Previously, security was centered on a single point of contact for all service demands. The resources are made available through many points of access that are interconnected to produce a unique solution in the microservice-based architecture. [4] Microservices are more difficult to implement than monolithic security services. Monolithic services have a distinct boundary that encloses their interconnections. This will mask security weaknesses in the system's deeper layers. A microservice's communications are also encapsulated. Microservices and services both have well-defined requirements.[2][3]

A simple routine completion in a microservice-based system, for example, requires the microservices to communicate with each other through the network. This increases the attack surface by exposing more data and information (endpoints) about the system. One of the primary issues with this tech-nique is the communication between other services in the same network, which requires some caution. Teams for the development of a system based on microservices are often separated into teams and services, with these teams being in charge of the construction and delivery of services. [3][5]

For this type of deployment, the teams must be aligned in terms of the microservices and their interconnection, as well as the protocol used to carry out the communication, in order to adhere to a standard for access protection and incorrect interception. The fundamental element of security is defining how services are interconnected and interact.[1]

The rising difficulty in debugging, monitoring, auditing, and forensic analysis of the entire program is a security risk brought on by such network complexity. Because microservices are frequently deployed in a cloud that the application owners do not control, constructing a global view of the complete program is difficult.

An application in microservice architecture is just a collection of workflows. These workflows can include multiple tiers of services, each of which processes and modifies data before it reaches its final destination. We require a method to validate metadata associated with a data stream and control its authenticity throughout time and re-elaboration.[2][5]

Many instances of a particular service may be running at any given time in a microservice architecture, and these instances may cease and start over time. The goal of service discovery is to make it possible for service consumers to find service providers in real time and communicate with them. Docker Containers have attracted a lot of attention due to their agility and ease of adding new services. The containers enable microservices to be packed and made available alongside their dependencies in a single image, allowing for faster service availability and reduced downtime. Code portability is the name for this mode.

In the context of microservices, the use of docker containers for service delivery has yielded benefits in terms of automation, independence, portability, and security, particularly when considering the docker platform's ease of management, creation, and continuous integration of environments systems. Each Docker container ideally contains only the application and the dependencies that the application requires to run, and no more or less.[2][3][5]

*B. Security Challenges for Cloud computing*

*1) Distributed Communication:* To communicate amongst participating microservices, CNA uses distributed communication mechanisms. These lightweight communication techniques, such as REST and Thrift, are relayed over the network, as is common of distributed systems. As a result, network attacks like MiTM and session/token hijacking can compromise microservice communications. A successful attack on these communication processes could have a detrimental influence on microservices collaboration and orchestration, as well as put the application's security at risk.

*2) Ephemeral Nature of Resource:* Microservices are deployed dynamically, which means that characteristics like IP addresses, port numbers, and service endpoints are always changing. Security procedures such as security assessments, which are generally configured for static network resources, hosts, and applications, are hampered by this dynamic. Methods for discovering microservice endpoints, identifying when microservice instances are scaled, and distinguishing versioned microservice instances all provide challenges to these security operations. This discoverability issue is similar to that of virtualized environments, but it happens at the application layer, therefore virtualization-based solutions can address it.

*3) Trust Amongst Inter-Communicating Microservices:* Assuming that all intercommunicating service instances run within the same security trust domain, certain security safeguards are relaxed. However, this may pose security risks; for example, an attacker who takes control of one microservice may be able to spread an attack to additional microservices.

*4) REST related Challenges:* For implementing CNA, REST is a popular design pattern. REST exposes resources through endpoints that can be accessed by a variety of applications, including mobile clients and IoT devices. However, unlike web applications, automated security assessments for RESTful applications are difficult to come by[4]. Iteratively retrieving and crawling through web-page links, or discovering entry and exit points, is how security scanners uncover vulnerabilities in web applications. Because web apps have well-defined interfaces, this is possible. Security scanners send random requests after the crawling phase and analyze responses for security vulnerabilities. Web services, on the other hand, do not have the same well-defined interfaces as web apps. As a result, finding entry and exit points is difficult for automated systems. Furthermore, unlike web apps, which create predictable results, RESTful applications' responses can be dynamically generated at request time.

*5) Cloud Specific Vulnerabilities:* CNA are deployed in cloud environments, making them vulnerable to the cloud's own security challenges, known as cloud-specific vulnerabilities [9]. The Cloud Security Shared Responsibility Model [3] specifically offered a method for overcoming these cloud-specific flaws. Application owners (cloud users) are responsible for protecting rented VMs or containers, establishing firewalls and security groups, and so on on Infrastructure as a Service (IaaS) platforms. CNA deployments are implicitly required to use this technique for securing deployed microservices.

*C. Ongoing Security Challenges for IoT*

*1) Heterogeneous distribution and interoperability:* Fine-grained, distributed, and independent entities make up microservices-based programs. These entities are built using various technologies, their data is kept in various databases, and they connect with one another via APIs that are independent of machine architecture and even programming language, hence increasing the overall number of potential security vulnerabilities.[8]

*2) Container's vulnerability:* Containers make developing, testing, deploying, and orchestrating microservices across multiple environments much easier. However, because they have various vulnerabilities, they represent an enticing target for hackers [31]. The data contained in the traffic generated by the communication required for container management and orchestration could be valuable to hackers. Furthermore, most of the primary components that make up a containerized system (such as hosts, registries, and images) can be targets for hackers. A hacked container can act as a gateway to other components of the IoT system, such as the system hosts, distant devices, and other interoperable containers, among other things.[8]

*3) Cloud Security threats:* IoT systems use loud-native computing as a development approach because it allows them to construct, run, and deploy distributed and scalable entities. However, this method raises a number of security risks [32]. With so much data entering the cloud, particularly public cloud services, hackers may be able to exploit or misuse it. Data breaches can result in significant damage to an organization's brand and confidence. They may also have an influence on short- and long-term revenues, as well as cause intellectual property loss and major legal obligations.[8]

*4) Authentication and Authorization:* Because if one service is hijacked by a hacker, the remaining services can be compromised and utilized maliciously, microservices-based applications should be able to check the validity of each service engaged in the communication. Furthermore, when a service receives a message, it must determine whether the message is suspicious and whether the source service is authorized to send it. To address these issues, a set of authorization and access control mechanisms (e.g., policies, models, and protocols) should be implemented to meet the security requirements of heterogeneous environments (e.g., multi-cloud environments) and to ensure centralized security management that effectively protects deployed applications.[8]

*5) Resource Limitations:* Because they are unable to execute complicated processing activities in real-time, most IoT devices are low-energy embedded devices that lack the computing resources required to facilitate the deployment of advanced and effective authentication and encryption techniques [6]. Dedicated security chips can be used to overcome this problem, but they increase cost, complexity, and energy consumption, thus manufacturers have little motivation to do so, especially in consumer-grade products. Because local storage is limited, IoT devices frequently rely on cloud storage to store the data they generate; this fact exposes IoT devices to extra threats in the form of cloud security concerns.[8]

V. CONCLUSION

I have surveyed the microservice architectural style in this paper, with a particular focus on the security implications. Microservices combine concepts from service orientation, distributed systems, and the abstraction, reuse, and separation of concerns principles of software engineering. This combination creates new security difficulties as well as old security challenges wrapped in a fresh package. The microservices model of highly separated, readily redeployable distributed components also suggests new security opportunities, such as increasing variety or limiting data access to only the services that require it.

Important features of software development using the microservice based architecture must be observed in order to achieve effective outcomes. The goal of this survey is also is to define the parts of microservices implementation in the cloud computing environment, explain and connect the elements that integrate standards and solutions relevant to microservices design.

I have also given the rundown of the current security threats affecting IoT setups. This study examines publications that propose the microservices-based development paradigm as a way to address some of the current security issues. It shows how microservices might improve security in such heterogeneous and distributed contexts using example studies from the recent literature.

In my mind, microservice and IoT have great potential to revolutionize the way we live, work, and interact in the future. Due to their scalability and flexibility, microservices are already being used to implement new applications related to machine learning, automation, data analytics, and more. Meanwhile, IoT offers numerous advantages such as decentralization [19-24], low-cost data transmission, safety control measures, real-time management of networks and devices, enhanced communication performance with reduced power consumption and parallel communication across protocols [25-34]. With a well-integrated microservice and IoT strategy in place, organizations can see vast improvements in efficiency and productivity. Given these benefits, it's no surprise microservices and IoT are becoming increasingly popular for a wide range of applications in various industries in the future [35-45].

REFERENCES

[1] Tetiana Yarygina,Anya Helene Bagge,"Overcoming Security Challenges in Microservice Architectures",(2018).
[2] Washington Henrique Carvalho Almeida, Luciano de Aguiar Monteiro, Raphael Rodrigues Hazin, Anderson Cavalcanti de Lima and Felipe Silva Ferraz, "Integrating Continuous Security Assessments in Microservices and Cloud Native Applications",(2017).
[3] Washington Henrique Carvalho Almeida, Luciano de Aguiar Monteiro, Raphael Rodrigues Hazin, Anderson Cavalcanti de Lima and Felipe

Silva Ferraz, "Survey on Microservice Architecture - Security, Privacy and Standardization on Cloud Computing Environment",(2017).
[4] Nuno Mateus-Coelho,Manuela Cruz-Cunha,Luis Gonzaga Ferreira, "Se- curity in Microservices Architectures",(2020).
[5] R.Divya mounika1, R.Naresh, "The concept of Privacy and Standardization of Microservice Architectures in cloud computing", (2020).
[6] Dongjin Yu, Yike Jin, Yuqun Zhang, Xi Zheng,"A survey on security issues in services communication of Microservices-enabled fog applica- tions",(2017) .
[7] Stephen Jacob, Yuansong Qiao ,and Brian Lee, "Detecting Cyber Security Attacks against a Microservices Application using Distributed Tracing" ,(2021).
[8] Maha Drissa, Daniah Hasanb, Wadii Boulila, Jawad Ahmad,"Microservices in IoT Security: Current Solutions, Research Challenges, and Future Directions",(2021).
[9] Mung Kim, Farokh B. Bastani, I-Ling Yen, Ing-Ray Chen,"High-Assurance Synthesis of Security Services from Basic Microservices",(2013).
[10] Yuqiong Sun, Susanta Nanda, Trent Jaeger, "Security-as-a-Service for Microservices-Based Cloud Applications", (2015).
[11] Marcela S. Melara, Mic Bowman ,"Enabling Security-Oriented Orches- tration of Microservices",(2021).
[12] "Analysing Privacy-Preserving Constraints in Microservices Architecture", Inna Vistbakka, Elena Troubitsyna,(2020).
[13] "Formalising Privacy-Preserving Constraints in Microservices Architec- ture", Inna Vistbakka, Elena Troubitsyna,(2020).
[14] Joe Chou,Eyhab Al-Masri, Sergey Kanzhelev, Hossam Fattah,"Detecting Security and Privacy Risks in Microservices End-to-End Communication Using Neural Networks",(2021).
[15] Ali Rezaei Nasab, Mojtaba Shahin, Seyed Ali Hoseyni Raviz, Peng Liang, Amir Mashmool, and Valentina Lenarduzzi,"An Empirical Study of Security Practices for Microservices Systems",(2021).
[16] Wesley K. G. Assunção, Thelma Elita Colanzi, Luiz Carvalho, Alessan- dro Garcia, Juliana Alves Pereira, Maria Julia de Lima, Carlos Lu- cena,"Analysis of a many-objective optimization approach for identi- fying microservices from legacy systems", (2022).
[17] Georgios L. Stavrinides, Helen D. Karatza,"Containerization, microser- vices and serverless cloud computing: Modeling and simulation", (2022).
[18] "Microservices in IoT Security: Current Solutions, Research Challenges, and Future Directions", Maha Drissa, Daniah Hasanb, Wadii Boulila, Jawad Ahmad,(2021).
[19] Kang K-D, Menasche DS, Kučuk G, Zhu T, Yi P. Edge computing in the Internet of Things. International Journal of Distributed Sensor Networks. 2017;13(9). doi:10.1177/1550147717732446
[20] Jianhang Liu, Shuqing Wang, Shibao Li, Xuerong Cui, Yan Pan, Ting Zhu, "MCTS: MultiChannel Transmission Simultaneously Using Non-Feedback Fountain Code," In IEEE Access, Pages 58373-58382, 2018.
[21] Zicheng Chi, Yan Li, Hongyu Sun, Yao Yao, Zheng Lu, and Ting Zhu, "B2W2 : N-Way Concurrent Communica- tion for IoT Devices," In the 14th ACM Conference on Embedded Networked Sensor System
[22] Zhichuan Huang, David Corrigan, Sandeep Narayanan, Ting Zhu, Elizabeth Bentley, and Michael Medley, "Dis- tributed and Dynamic Spectrum Management in Airborne Networks", IEEE Military Communications Conference 2015, Tampa, FL, October, 2015.
[23] Zhu, Ting, Qingquan Zhang, Sheng Xiao, Yu Gu, Ping Yi, and Yanhua Li. "Big data in future sensing." Interna- tional Journal of Distributed Sensor Networks (2015).
[24] Liang He, Linghe Kong, Yu Gu, Jianping Pan, and Ting Zhu, "Evaluating the On-Demand Mobile Charging in Wireless Sensor Networks," In IEEE Transactions on Mobile Computing, 2014.
[25] Bhaskar AV, Baingane A, Jahnige R, Zhang Q, Zhu T. A Secured Protocol for IoT Networks. arXiv preprint arXiv:2012.11072. 2020 Dec 21.
[26] Vishnu Bhaskar A, Baingane A, Jahnige R, Zhang Q, Zhu T. A Secured Protocol for IoT Networks. arXiv e-prints. 2020 Dec:arXiv-2012.
[27] Z. Chi, Y. Li, Y. Yao and T. Zhu, "PMC: Parallel multi- protocol communication to heterogeneous IoT radios within a single WiFi channel," 2017 IEEE 25th Interna- tional Conference on Network Protocols (ICNP), 2017, pp. 1-10, doi: 10.1109/ICNP.2017.8117550.
[28] F. Chai, T. Zhu and K. -D. Kang, "A link-correlation- aware cross-layer protocol for IoT devices," 2016 IEEE International Conference on Communications (ICC), 2016, pp. 1-6, doi: 10.1109/ICC.2016.7510940.
[29] Zhou, Z., Xie, M., Zhu, T., Xu, W., Yi, P., Huang, Z., Zhang, Q. and Xiao, S., 2014, November. EEP2P: An energy-efficient and economy-efficient P2P network protocol. In International Green Computing Conference (pp. 1-6). IEEE.
[30] Xie M, Zhou Z, Zhu T, Xu W, Yi P, Huang Z, Zhang Q, Xiao S. EEP2P: An Energy-Efficient and Economy- Efficient P2P Network Protocol. IEEE/CIC ICCC. 2014.
[31] Ting Zhu and Ping Yi, " Chapter 17: Reliable and Energy- Efficient Networking Protocol Design in Wireless Sensor Networks, " In Intelligent Sensor Networks: The Integra- tion of Sensor Networks, Signal Processing and Machine Learning, Taylor & Francis LLC, CRC Press, December 2012.
[32] Z. Chang and Z. Ting, " Thorough Analysis of MAC Protocols in Wireless Sensor Networks, " 2008 4th In- ternational Conference on Wireless Communications, Networking and Mobile Computing, 2008, pp. 1-4, doi: 10.1109/WiCom.2008.887.
[33] T. Zhu and M. Yu, " A Dynamic Secure QoS Routing Protocol for Wireless Ad Hoc Networks," 2006 IEEE Sarnoff Symposium, 2006, pp. 1-4, doi: 10.1109/SARNOF.2006.4534791.
[34] T. Zhu and M. Yu, "NIS02-4: A Secure Quality of Ser- vice Routing Protocol for Wireless Ad Hoc Networks," IEEE Globecom 2006, 2006, pp. 1-6, doi: 10.1109/GLO- COM.2006.270.
[35] Ting Zhu, Aditya Mishra, David Irwin, Navin Sharma, Prashant Shenoy, and Don Towsley. 2011. The case for efficient renewable energy management in smart homes. In Proceedings of the Third ACM Workshop on Embedded Sensing Systems for Energy- Efficiency in Buildings (BuildSys '11). Association for Computing Machinery, New York, NY, USA, 67–72. https://doi.org/10.1145/2434020.2434042
[36] Ting Zhu, and Chang Zhou, " An Efficient Renewable Energy Management and Sharing System for Sustainable Embedded Devices, " In Journal of Electrical and Computer Engineering, Volume 2012, doi: 10.1155/2012/185959.
[37] P. Yi, T. Zhu, B. Jiang, B. Wang and D. Towsley, "An energy transmission and distribution network using electric vehicles," 2012 IEEE International Conference on Communications (ICC), 2012, pp. 3335-3339, doi: 10.1109/ICC.2012.6364483.
[38] Ting Zhu, Zhichuan Huang, Ankur Sharma, Jikui Su, David Irwin, Aditya Mishra, Daniel Menasche, and Prashant Shenoy, "Sharing Renewable Energy in Smart Microgrids," In the ACM/IEEE 4th International Conference on Cyber-Physical Systems (ACM/IEEE ICCPS '13), Philadelphia, PA, April 2013. (Acceptance rate: 23%)
[39] Aditya Kr Mishra, David Irwin, Prashant Shenoy, and Ting Zhu, "Scaling Distributed Energy Storage for Grid Peak Reduction," In the Fourth International Conference on Future Energy Systems (ACM e-Energy 2013), Berkeley, CA, 2013. (Acceptance rate: 28.9%, finalist for the Best Paper Award)
[40] Weigang Zhong, Zhichuan Huang, Ting Zhu, Yu Gu, Qingquan Zhang, Ping Yi, Dingde Jiang and Sheng Xiao. "iDES: Incentive- Driven Distributed Energy Sharing in Sustainable Microgrids", In 2014 IEEE International Green Computing Conference (IEEE IGCC '14), Dallas, TX, November, 2014.
[41] Yunlong Hu, Ping Yi, Yu Sui, Zongrui Zhang, Yao Yao, Wei Wang, and Ting Zhu, "Dispatching and Distributing Energy in Energy Internet under Energy Dilemma," In 2018 Third International Conference on Security of Smart Cities, Industrial Control System and Communications (IEEE SSIC '18)..
[42] Bhaskar AV, Baingane A, Jahnige R, Zhang Q, Zhu T. A Secured Protocol for IoT Networks. arXiv preprint arXiv:2012.11072. 2020 Dec 21.
[43] Zicheng Chi, Yan Li, Xin Liu, Yao Yao, Yanchao Zhang, and Ting Zhu. 2019. Parallel inclusive communication for connecting heterogeneous IoT devices at the edge. In Proceedings of the 17th Conference on Embedded Networked Sensor Systems (SenSys '19). Association for Computing Machinery, New York, NY, USA, 205–218. https://doi.org/10.1145/3356250.3360046
[44] Yan Pan, Shining Li, Bingqi Li, B. Bhargav, Zeyu Ning, and Ting Zhu, "When UAVs coexist with manned airplanes: Large-scale aerial network management using ADS-B," In Transactions on Emerging Telecommunications Technologies, Volume 30, Issue 10, 2019. https://doi.org/10.1002/ett.3714
[45] W. Wang, X. Liu, Y. Yao, Y. Pan, Z. Chi and T. Zhu, "CRF: Coexistent Routing and Flooding using WiFi Packets in Heterogeneous IoT Networks," IEEE INFOCOM 2019 - IEEE Conference on Computer Communications, 2019, pp. 19-27, doi: 10.1109/INFOCOM.2019.8737525.